\pdfoutput=1 

\documentclass[a4paper,11pt]{article}

\usepackage{lipsum}
\usepackage{layout}

\newcommand\blfootnote[1]{%
  \begingroup
  \renewcommand\thefootnote{}\footnote{#1}%
  \addtocounter{footnote}{-1}%
  \endgroup
}

\usepackage{verbatim}
\usepackage{cite}
\usepackage{graphicx}
\usepackage{float}

\begin{document}

\title{\textbf{DMAPS: a fully depleted monolithic active pixel sensor - analog performance characterization}}

\author{M. Havr\'anek\textsuperscript{*, **}, T. Hemperek, H.  Kr\"uger, Y. Fu, L. Germic,\\
T.~Kishishita, T.~Obermann and~N. Wermes}
\date{} 
\maketitle

\begin{center}
\textit{Physikalisches Institut,  Universit\"at Bonn\\
 Nu{\ss}allee 12, 53115 Bonn, Germany \textsuperscript{\(\dagger\)}}

\end{center}
\thispagestyle{empty}

\blfootnote{\hspace{-5mm}\textsuperscript{*} Corresponding author\\
\indent Email: miroslav.havranek@cern.ch}
\blfootnote{\hspace{-5mm}\textsuperscript{**}   Now at FNSPE, Czech Technical University, B\v{r}ehov\'a 7, 115 19 Praha, Czech Republic}
\blfootnote{\hspace{-5mm}\textsuperscript{\(\dagger\)} Work supported by the German Deutsche Forschungsgemeinschaft DFG under contract WE 976/4-1}

\vspace{2cm}

\textbf{Abstract}\\

Monolithic Active Pixel Sensors (MAPS) have been developed since the late 1990s  based on silicon substrates 
with a thin epitaxial layer (thickness of 10-15~$\mu$m) in which charge is~collected on an electrode, albeit by disordered and 
slow diffusion rather than by drift in a directed electric field. As a consequence, the signal is small ($\approx$ 1000 e\textsuperscript{-}) 
and the radiation tolerance is much below the LHC requirements by factors of 100 to 1000. In this paper we present the development of a~fully 
Depleted Monolithic Active Pixel Sensors (DMAPS) based on a high resistivity substrate allowing~the~creation of a fully depleted 
detection volume. This concept overcomes the inherent limitations of charge collection by diffusion in the standard MAPS designs. 
We present results from a test chip EPCB01 designed in~a~commercial 150~nm CMOS technology. The technology provides a thin 
(\(\sim\) 50 \(\mu\)m) high resistivity n-type silicon substrate as well as an additional deep p-well which allows to integrate full CMOS 
circuitry inside the pixel. Different matrix types with several variants of collection electrodes have been implemented. Measurements 
of the analog performance of this first implementation of DMAPS pixels will be presented.\\

\vspace{1mm}


\vspace{5mm}


\newpage

\section{Introduction}\label{sec:intro}
Monolithic Active Pixel Sensors (MAPS) have been proposed~\cite{meynants},~\cite{turch} as tracking detectors for particle physics experiments allowing high spatial 
resolution and a superior material budget compared to hybrid pixel detectors~\cite{wermes}. Unlike the latter, MAPS integrate both sensor and front-end (FE) electronics 
in a single silicon chip and thus do not require costly and often difficult chip-to-sensor interconnection. 
However, standard MAPS pixel sensors have two drawbacks which limit their area of application: First, in a standard CMOS process the common implant well configuration 
does not allow to use both transistor types (NMOS and PMOS) in the pixel area, thus severely limiting the~complexity of the electronics circuitry. 
To overcome this limitation, an~additional implant must be added in the technology to isolate the electronics from the charge collection nodes as introduced for example 
in \cite{quad_well} and \cite{cherwell}. Second,  the collection of charge released from a particle or radiation is accomplished by diffusion rather than by drift in an electric 
field and is hence slow, severely limiting the achievable time resolution (e.g. for time stamping) and rendering the sensor more vulnerable to bulk damage by non-ionizing 
radiation especially to levels important at LHC~\cite{atlas},~\cite{cms}. The low resistivity of the Si substrate in standard CMOS processes (typically about 
10 \(\Omega \cdot\)cm) only allows the creation of a very small depletion layer formed around the charge collecting p-n junction. Attempts have been 
made~\cite{strasbourg} using epitaxial layers of a few~$\mu$m high purity silicon to allow more charge to be collected. Some CMOS technologies allow 
using high voltage through which the silicon substrate can be depleted by $\sim10~\mu$m \cite{peric},  but the rest of the silicon remains undepleted.
In this paper we report on an implementation of an advanced MAPS concept employing full CMOS (NMOS and PMOS) 
electronics in the active area of pixels using a high resistivity (detector-grade) silicon substrate depleted by a bias voltage. 
The technology is introduced and the~performance of a~test chip EPCB01 which contains different 
circuit variants is characterized with respect to analog performance and tolerance to ionizing radiation.


\section{Test chip EPCB01}\label{sec:test_chip}

\begin{figure}[h]
\centering
\begin{minipage}[c]{420pt}
\includegraphics[width=420pt]{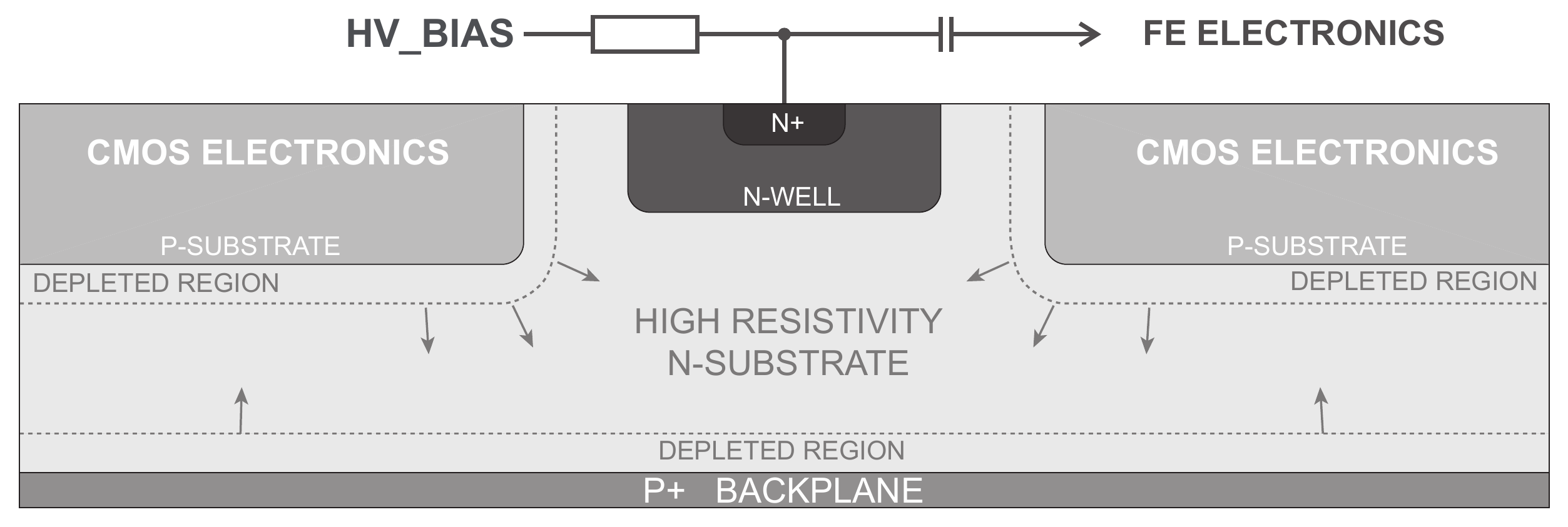}
\end{minipage}

\begin{minipage}[t]{420pt}
\caption{Cross section of the DMAPS pixel.}
\label{fig:epcb_cross_sec}
\end{minipage}
\end{figure}

To investigate the concept of the DMAPS pixels and their applicability in high energy physics,~a~prototype chip - EPCB01 
has been designed, fabricated and tested. The fabrication process is a~150~nm CMOS process using a high resistivity n-type silicon substrate. 
The chip itself has been thinned down to a thickness of 50~\(\mu\)m. A schematic cross section of a DMAPS pixel is shown in figure~\ref{fig:epcb_cross_sec}. \\  

A deep p-well is implanted in the n-type high resistivity substrate forming a p-substrate for the~integration of the CMOS electronics.
Transistors are contained within deep n-well which allows setting of the p-substrate potential independently of CMOS electronics.   
The sensitive elements are implemented by n-wells implanted in the high resistivity~n-type substrate. Although these n-wells (charge collection electrodes) do not form p-n junctions, 
they create regions with high electric potential and thus build up an electric drift field in the depleted bulk.
The depletion region is formed between the p-substrate and the~high resistivity n-substrate and between the p+ backplane and~the~high resitivity n-substrate 
(see figure~\ref{fig:epcb_cross_sec}).
This technology allows biasing of the~charge collecting n-wells with high voltage (tested up to 15 V), while the FE electronics operates at 1.8~V. 
More information about the technology can be found in~\cite{espros}. The collection electrode collects electrons, which are the preferred charge carriers due 
to their higher mobility with respect to holes. In theory, if full depletion is achieved, a signal of approximatelly 4000~e\textsuperscript{-} per minimum ionizing 
particle (MIP) can be collected in this sensor configuration. Advantages of this type of~sensor are therefore apparent: large signal, fast signal collection and small 
sensor capacitance (potentially low noise). Six different pixel arrays with dimensions of 8\(\times\)8 and 6\(\times\)8~pixels have been integrated in the 
EPCB01. These arrays differ by architecture of~the FE electronics, biasing, coupling and geometry of~the~sensitive elements. \\


\section{DMAPS pixel}\label{sec:pixel}

\begin{figure}[H]
\centering
\begin{minipage}[c]{420pt}
\includegraphics[width=420pt]{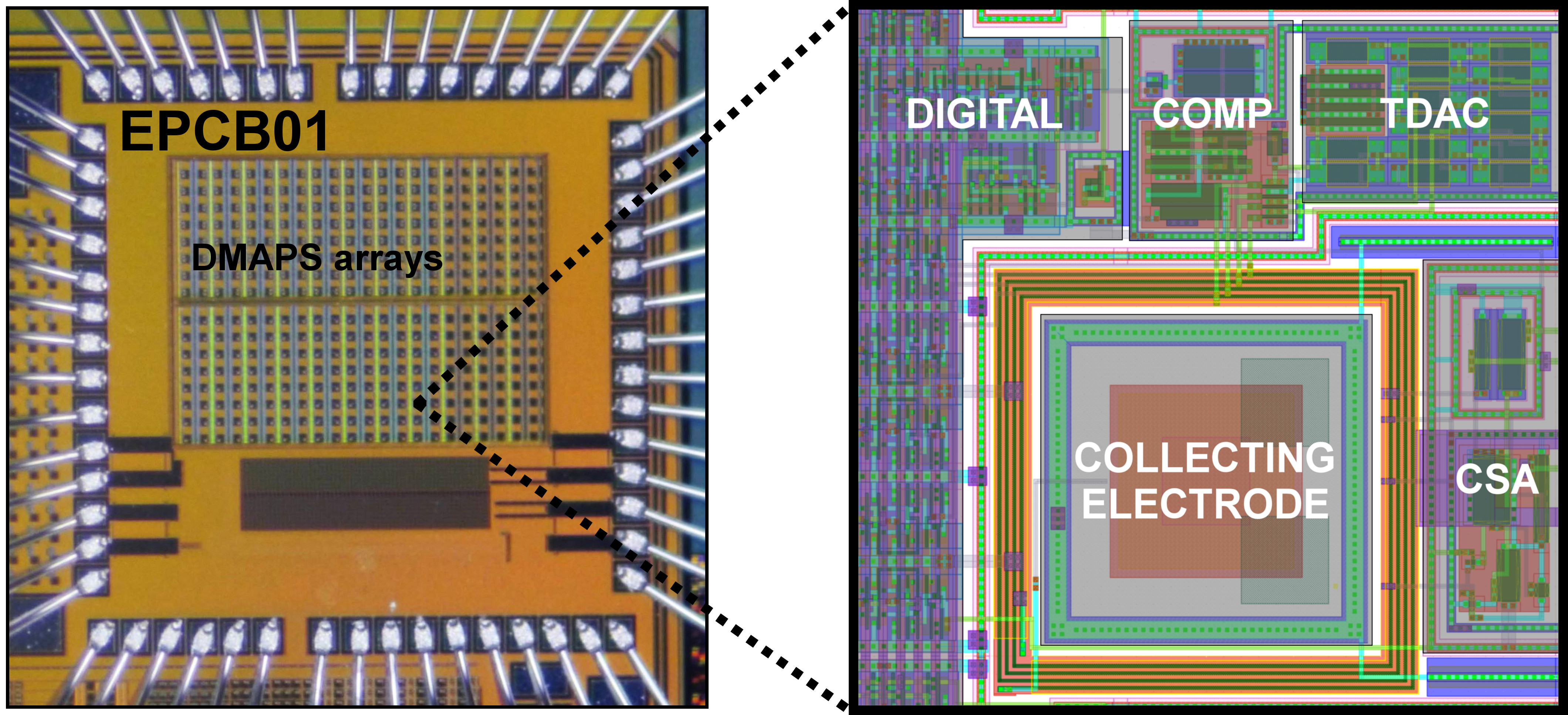}
\end{minipage}

\begin{minipage}[t]{420pt}
\caption{Micrograph of EPCB01 and layout of a single DMAPS pixel.}
\label{fig:pixel}
\end{minipage}
\end{figure}

A single DMAPS pixel implemented in the EPCB01 can be divided in three sections: 1. high-resistive collection part biased with high voltage, 2. analog FE electronics and 
3. digital logic for pixel configuration and read-out. Micrograph of EPCB01 and layout of one DMAPS pixel with highlighted functional blocks is shown in figure~\ref{fig:pixel}.
The pixel size is 40\(\times\)40 \(\mu\)m\textsuperscript{2} and the~sensitive part occupies 25\% of the~total pixel area.~The circuitry inside a pixel contains about 
160-180 transistors (depending on the~variant, see table~\ref{tab:versions}).

\begin{figure}[h]
\centering
\begin{minipage}[c]{420pt}
\includegraphics[width=420pt]{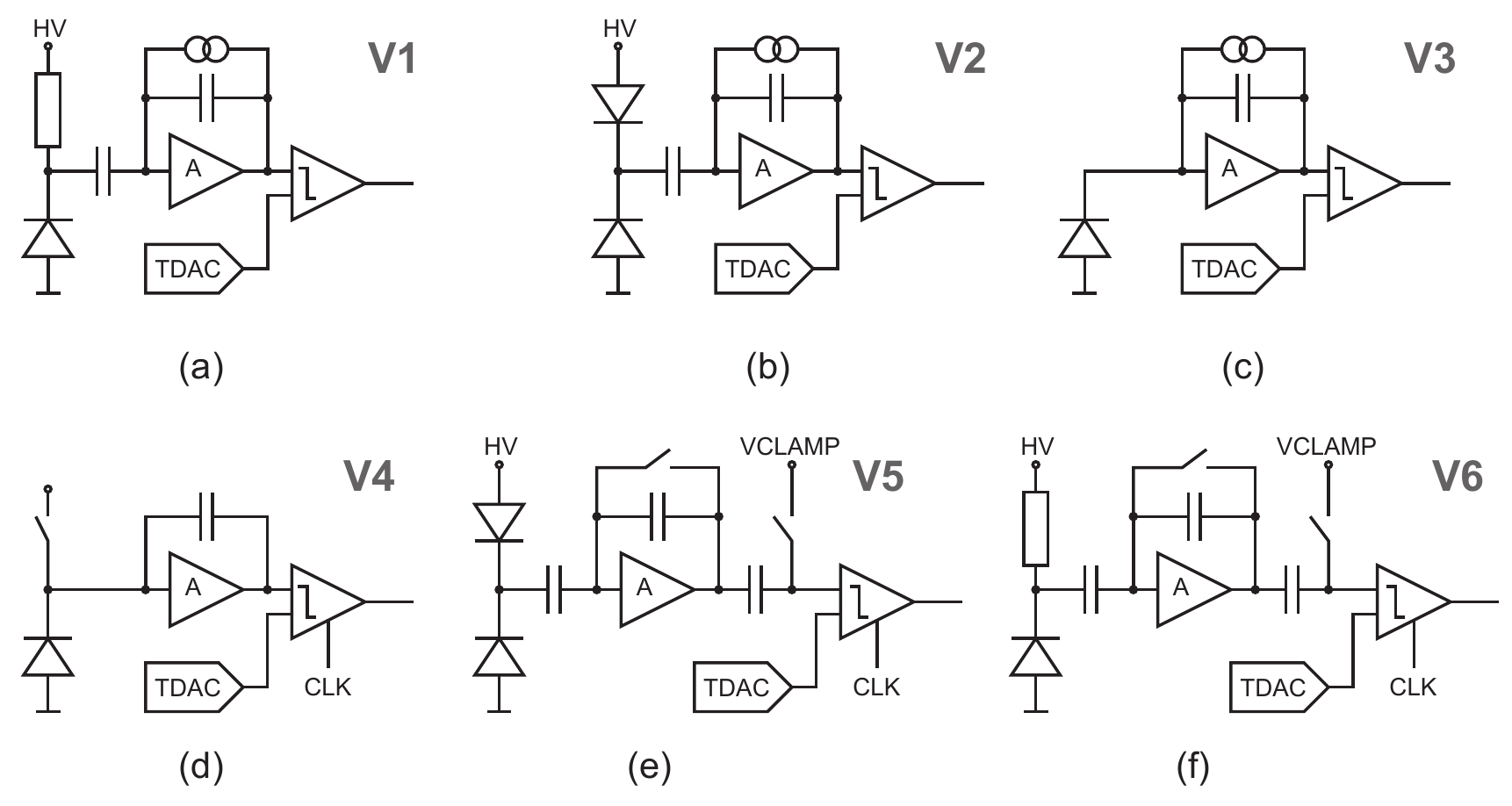}
\end{minipage}

\begin{minipage}[t]{420pt}
\caption{Six different variants of FE electronics V1-V6 shown in figures (a-f) are implemented in EPCB01.}
\label{fig:versions_all}
\end{minipage}
\end{figure}

The analog pixel FE electronics follows a similar scheme as the FE electronics commonly used in hybrid pixel detectors. The analog part contains a Charge Sensitive Amplifier (CSA),~a~discriminator with a DAC (TDAC) for threshold equalization. Two different architectures of the analog FE electronics have been implemented - (time)~continuous and switched. The continuous architecture uses a CSA with a continuous current source feedback followed by standard (time-continuous) 
discriminator. The~switched architecture uses a~CSA with a switched feedback and a~clamp-and-sample circuit followed by a dynamic discriminator. Motivation for 
implementation of the~clamp-and-sample circuit is reduction of the low frequency component of the noise. Resetting of the~CSA,
clamping and sampling is synchronous with a clock frequency of 1 MHz in contrast to the~continuous variant.
The digital part stores the configuration data for the analog part and enables the~read-out of~the~pixel matrix. The pixel has binary resolution and read-out is implemented 
through a~shift register. Different variants of the DMAPS pixel matrices are labeled V1-V6 and they are described~in table~\ref{tab:versions}.
A simplified schematic of each variant is shown in figure~\ref{fig:versions_all}.\\

\begin{table}[h]
\begin{center}
\begin{tabular}{|c|c|c|c|c|}
  \hline
  \hline Pixel variant & Biasing & Coupling & FE architecture & Matrix dimensions\\
  \hline
  \hline V1 & resistor & AC & continuous & 8 \(\times\) 8\\
  \hline V2 & diode & AC & continuous & 8 \(\times\) 8\\
  \hline V3 &  CSA feedback & DC & continuous & 6 \(\times\) 8\\
  \hline V4 &  switched  & DC & switched & 6 \(\times\) 8\\
  \hline V5 & diode & AC & switched & 8 \(\times\) 8\\
  \hline V6 & resistor & AC & switched & 8 \(\times\) 8\\
  \hline
\end{tabular}	
\caption {Different variants of the DMAPS pixel matrices implemented in EPCB01.}
\label{tab:versions}
\end{center}
\end{table}

\section{Performance of the EPCB01 prototype chip}\label{sec:analog_perf}

The (analog) performance of the DMAPS pixels is studied by an external charge injection with an~in-pixel 2 fF injection 
capacitor as well as with radioactive sources \textsuperscript{90}Sr and \textsuperscript{55}Fe.

\subsection{Gain determination with charge injection}  
 
\begin{figure}[h]
\centering
\begin{minipage}[c]{300pt}
\includegraphics[width=300pt]{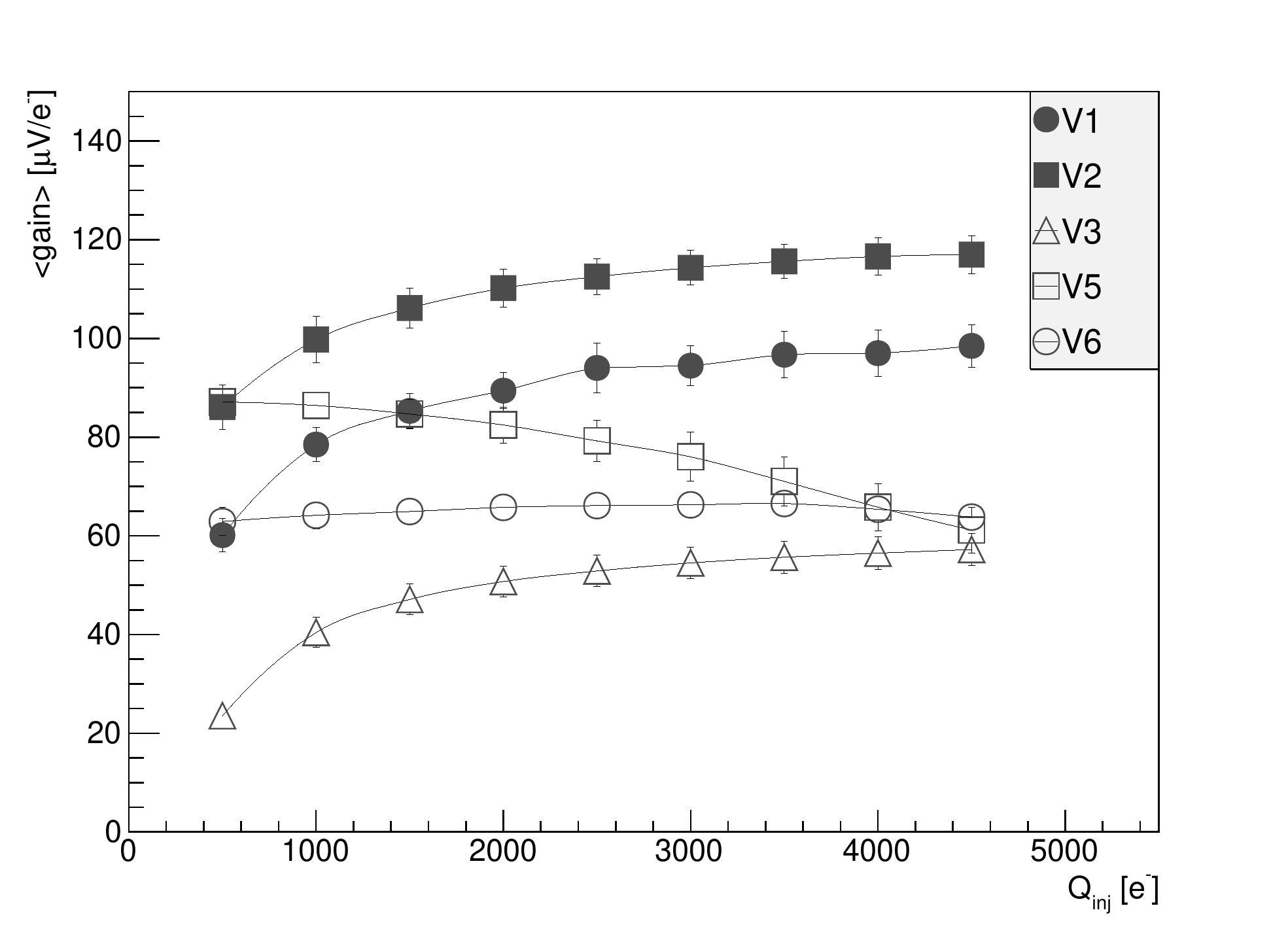}
\end{minipage}

\begin{minipage}[t]{300pt}
\caption{Mean gain of the FE electronics integrated in~DMAPS pixels. Size of the error bars has been scaled down by factor of~2.}
\label{fig:gain_mean_all}
\end{minipage}
\end{figure}

All pixel matrices except V4, which does not contain an injection capacitor, have been examined by external charge injection.
A high bias voltage of 11~V has been connected to the collection electrodes, keeping them depleted during the measurements. P-substrate and the chip backplane
have been biased with small negative voltage of -1.5~V. The best analog performance has been achieved with the matrix V2, which will therefore be the design variant on which 
we focus for the~rest of the~paper. In general, the pixel variants with a continuous rather than a switching circuit architecture have better performance 
than the switched variants. The switched variants suffer from a large "kick-back" caused by~the~dynamic discriminator. When the discriminator 
samples the signal, the~clamping capacitor is discharged by a large amount of parasitic charge injected from the~discriminator. 
This discharge is non-linear. It depends on the~signal amplitude. In addition, the~clamping capacitor is implemented by a MOSCAP capacitor, 
which is non-linear as well. Figure~\ref{fig:gain_mean_all} shows the gain of the FE electronics as a function of the injected electric charge. Each point represents a mean gain of 
all pixels of the particular pixel matrix and its error bar represents a standard deviation of the gain arising from gain variation from pixel to pixel.  
The error bars have been down-sized by a factor of 2 to keep the graph readable.

\begin{figure}[h]
\centering
\begin{minipage}[c]{300pt}
\includegraphics[width=300pt]{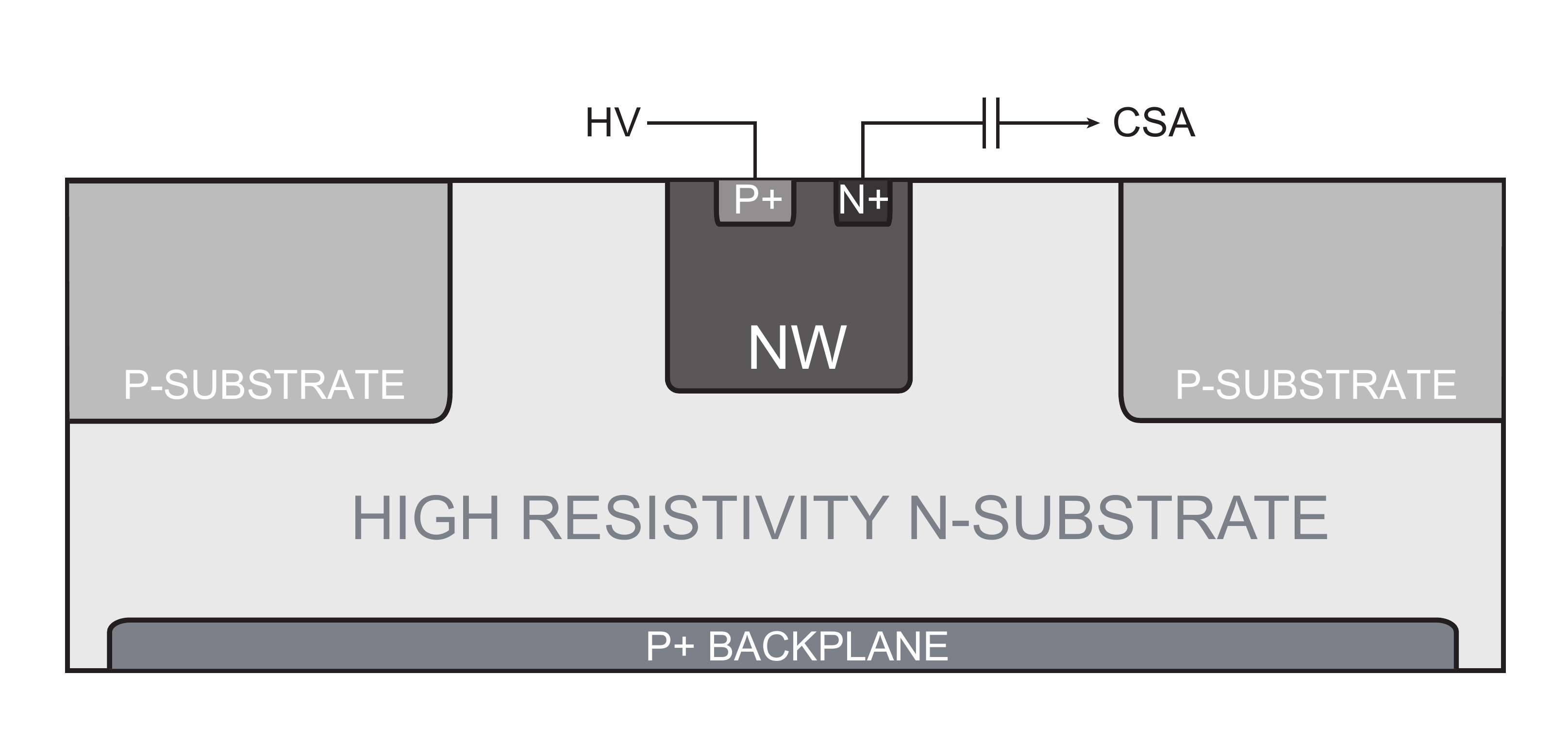}
\end{minipage}

\begin{minipage}[t]{300pt}
\caption{Cross section of the collection electrode used with diode biased pixel variants (V2 and V5).}
\label{fig:sensors}
\end{minipage}
\end{figure}

The highest gain of 99.8 \(\mu\)V/e\textsuperscript{-} measured after the injection of 1 ke\textsuperscript{-} (typical operating threshold) has been achieved with the pixel matrix V2.
The gain dispersion due to electronic mismatch within matrix variant V2 is 19.1~\(\mu\)V/e\textsuperscript{-}. Particularly interesting is the comparison of the matrices V1 and V2. 
The FE electronics in both matrices is identical (including the layout). The~shape of~the~gain dependence on the injected charge (gain curve) is similar in both cases, but they are 
shifted by 20.5~\(\mu\)V/e\textsuperscript{-} on average. The~gain shift most likely emerges due to the different capacitances of the~collection electrodes used in these pixel variants.
A~schematic cross section of the~collection electrode used with diode biased variants is shown in figure~\ref{fig:sensors}. Design of the collection electrode used with resistor biasing
has been provided by the foundry after design submission and we do not have access to the exact layout parameters. However, if the design of this collection electrode contains wider
n-well or n+ diffusion with respect to the collection electrode using diode biasing, this effect may easily increase capacitance of the collection electrode. The~CSA uses a~common 
source stage with a relatively small open loop gain of about 76 in the all pixel variants. 
Therefore the closed loop gain is sensitive to the sensor capacitance. Even greater shift in gain has been observed in case of V3 with respect to V2.~The~collection electrode in V3 is DC coupled to the FE electronics and is biased by a voltage of about 370~mV provided by CSA feedback. The~collection electrode is not fully depleted at this voltage. Therefore the capacitance 
is even higher than in V1 and~the~gain is therefore smaller. Almost the~same shifts of the gain-curves have been observed in case of V5 and V6 (at the beginning of their dynamic range). 
However, the~character of~the~non-linearity of the switched variants is different with respect to the continuous variants.~The~gain-curves of V5 and V6 are not equidistant and the~reason for 
this behavior is not clear. 

\subsection{Noise performance}

\begin{figure}[H]
\centering
\begin{minipage}[c]{300pt}
\includegraphics[width=300pt]{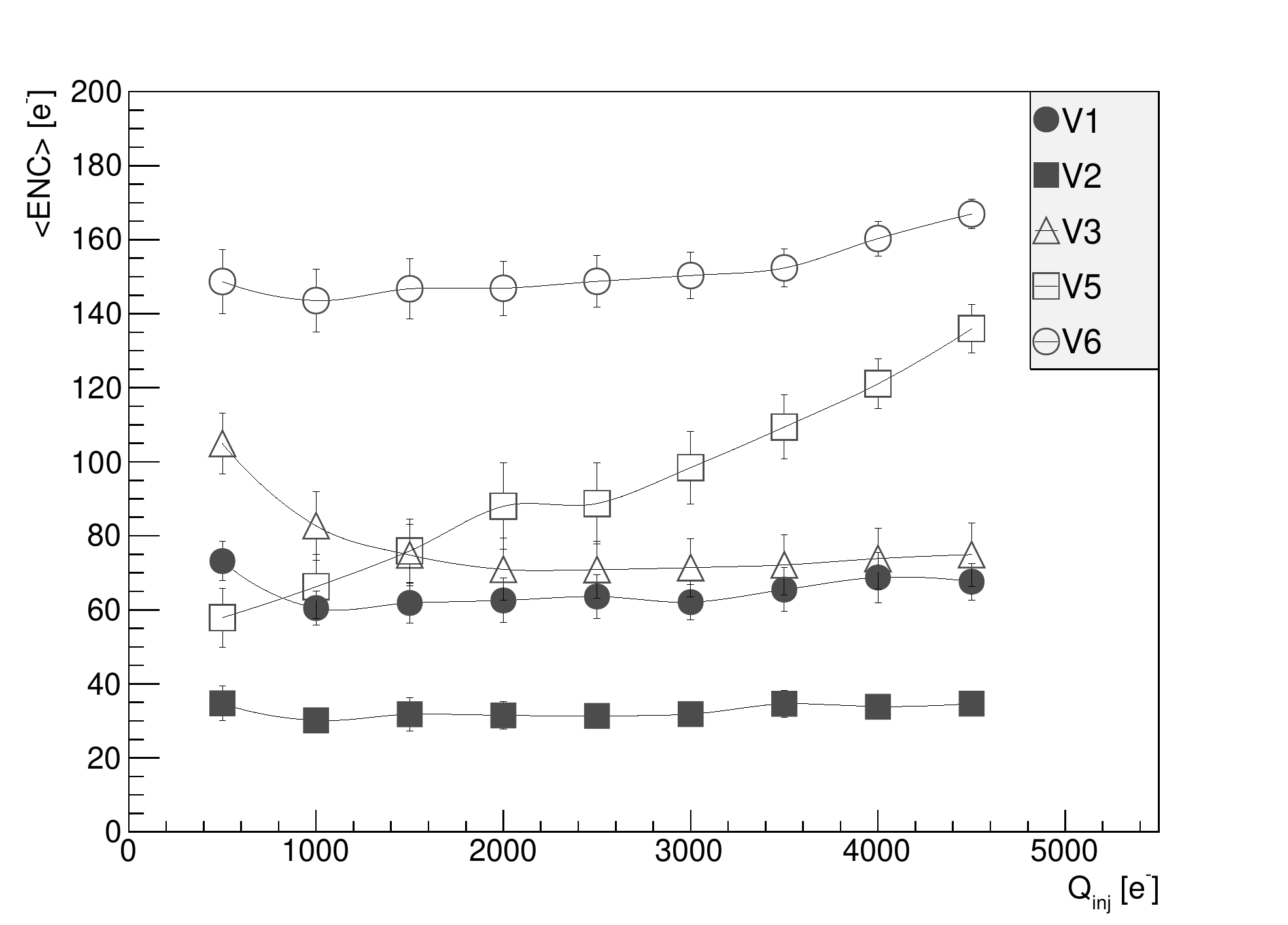}
\end{minipage}

\begin{minipage}[t]{300pt}
\caption{Mean ENC of the FE electronics integrated in the~DMAPS pixels. Size of the error bars has been scaled down by a factor of~2.}
\label{fig:enc_mean_all}
\end{minipage}
\end{figure}

The noise performance of the DMAPS pixels has been determined by threshold scans and s-curve fits and evaluated in terms of Equivalent Noise Charge (ENC). 
ENC as a function of injected charge is displayed in figure~\ref{fig:enc_mean_all}.

\begin{figure}[H]
\centering
\begin{minipage}[c]{300pt}
\includegraphics[width=300pt]{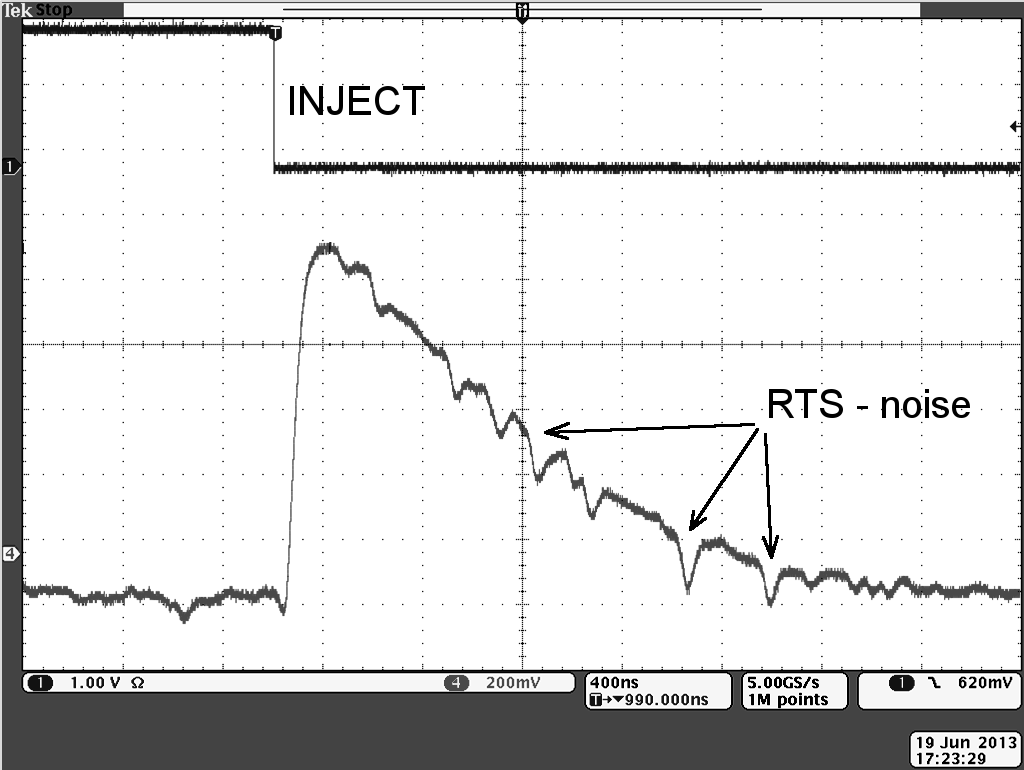}
\end{minipage}

\begin{minipage}[t]{300pt}
\caption{RTS noise observed at the analog output of the DMAPS pixel.}
\label{fig:rts}
\end{minipage}
\end{figure}

The lowest noise has been measured with the 
matrix V2. The differences of the ENC-curves of the matrices V1, V2 and V3 can be attributed to~the~differences of the~sensor capacitance. 
Each of the switched variants behaves differently in terms of noise. For small signals, the noise of V5 is comparable with its continuous 
counterpart (V2) and increases with increasing signal, because at the same time the gain decreases. The noise of variant V6 is much higher than in case of V5.
Random Telegraph Signal noise (RTS) represents a significant noise component at all variants of the DMAPS pixels in EPCB01. 
Spikes of the RTS noise superimposed on the signal from a DMAPS pixel si shown in figure~\ref{fig:rts}. 
RTS noise is not speciality of EPCB01 only. RTS noise often appears 
in monolithic pixels as described for example in~\cite{rts1}, \cite{rts2} and certain precautions have to be made in the design to minimize
its effect.

\subsection{Threshold dispersion}
\begin{figure}[h]
\centering
\begin{minipage}[c]{420pt}
\includegraphics[width=420pt]{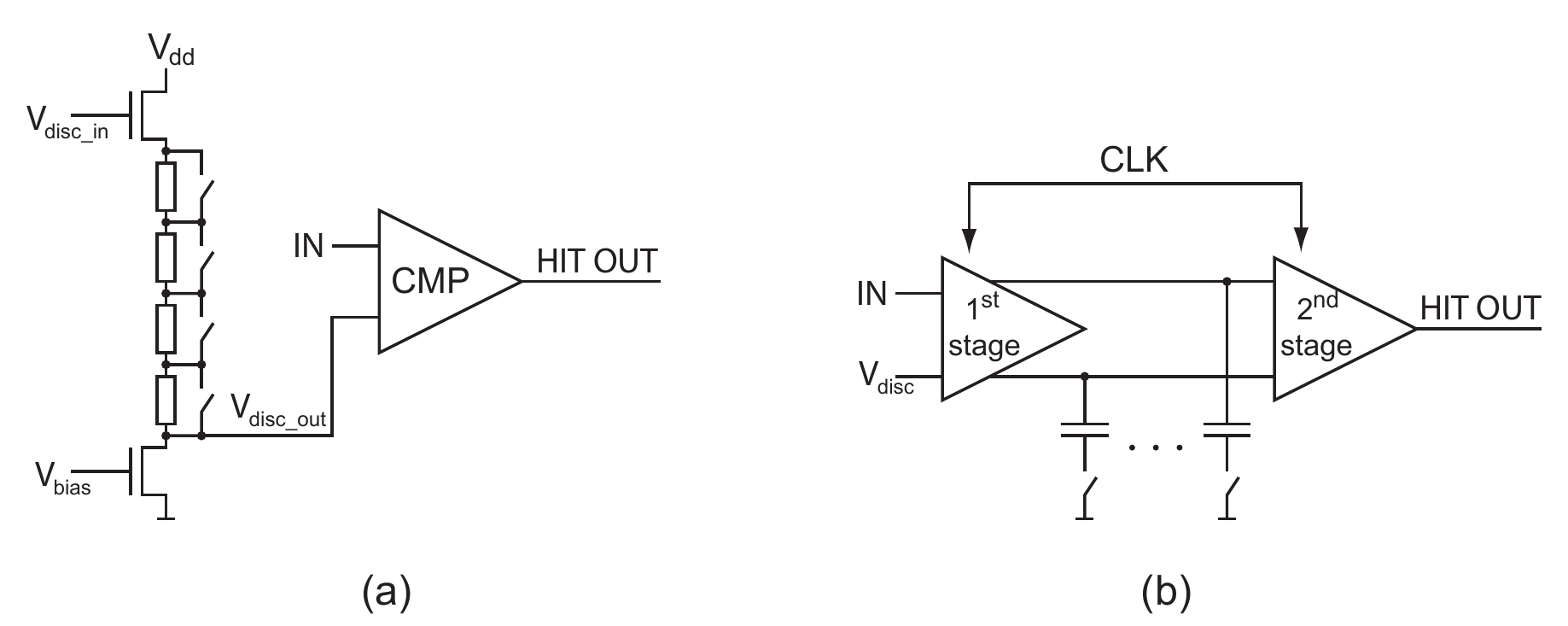}
\end{minipage}

\begin{minipage}[t]{420pt}
\caption{Two different variants of a discriminator with TDAC have been used in EPCB01. A continuous discriminator with resistive TDAC (a) and dynamic discriminator with 
switched capacitors TDAC (b).}
\label{fig:tdacs}
\end{minipage}
\end{figure}

Applications in high energy physics often require a uniform detection threshold across~the~pixel matrix. Dispersion of the threshold 
originates from the fabrication process variations and mismatch of the integrated electronic components. 
Two effects are responsible for the threshold dispersion. The~first is the dispersion of the quiescent voltage at the 
output of the CSA (baseline dispersion) and the second effect is the dispersion of the voltage offset of the discriminator. Each DMAPS pixel
contains a 4-bit DAC (TDAC) allowing threshold equalization.~The~implementation of the~TDAC is different in the continuous variants (V1, V2 and V3)
and in the~switched variants (V4, V5 and V6). The~continuous variants use a resistive TDAC embedded in a source follower as shown in figure~\ref{fig:tdacs}~(a). 
The~advantage of this solution is the possibility to adjust the threshold tuning range if needed. However, the output voltage of the TDAC is not a linear function of the input voltage,
resistors occupy a large area in the pixel and the entire TDAC consumes an additional power.~The~switched variants use a two stage dynamic discriminator. 
The input voltage offset of the~dynamic discriminator is very sensitive to the capacitance of the routing.  
This fact has been used in the design of a switched capacitance TDAC as shown in figure~\ref{fig:tdacs}~(b). By adjusting the capacitance between two 
branches of the discriminator, the voltage offset (discriminating threshold) can be adjusted. This TDAC does not introduce non-linearity to the 
threshold setting. In addition, this TDAC is very compact in the~pixel layout and does not increase the power budget of the pixel.

\begin{figure}[h]
\centering
\begin{minipage}[c]{420pt}
\includegraphics[width=420pt]{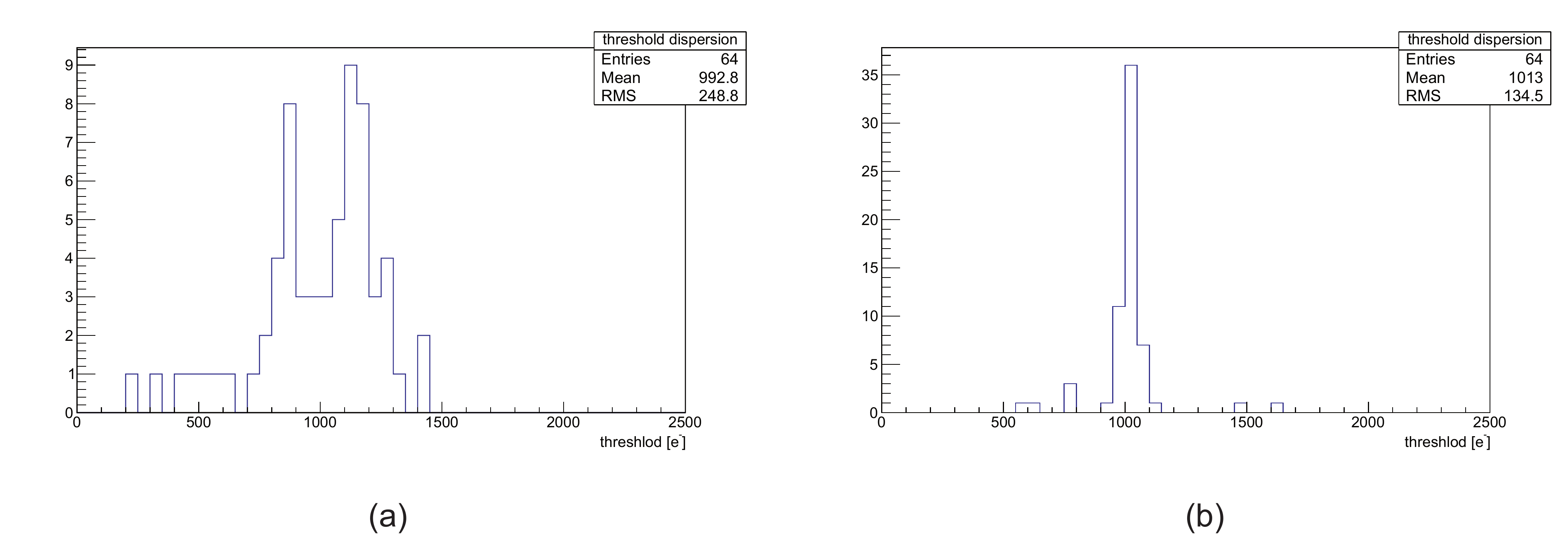}
\end{minipage}

\begin{minipage}[t]{420pt}
\caption{Threshold dispersion of the DMAPS pixels of variant V2 before (a) and after threshold tuning (b).}
\label{fig:thr_disp_v2}
\end{minipage}
\end{figure}

\begin{figure}[H]
\centering
\begin{minipage}[c]{420pt}
\includegraphics[width=420pt]{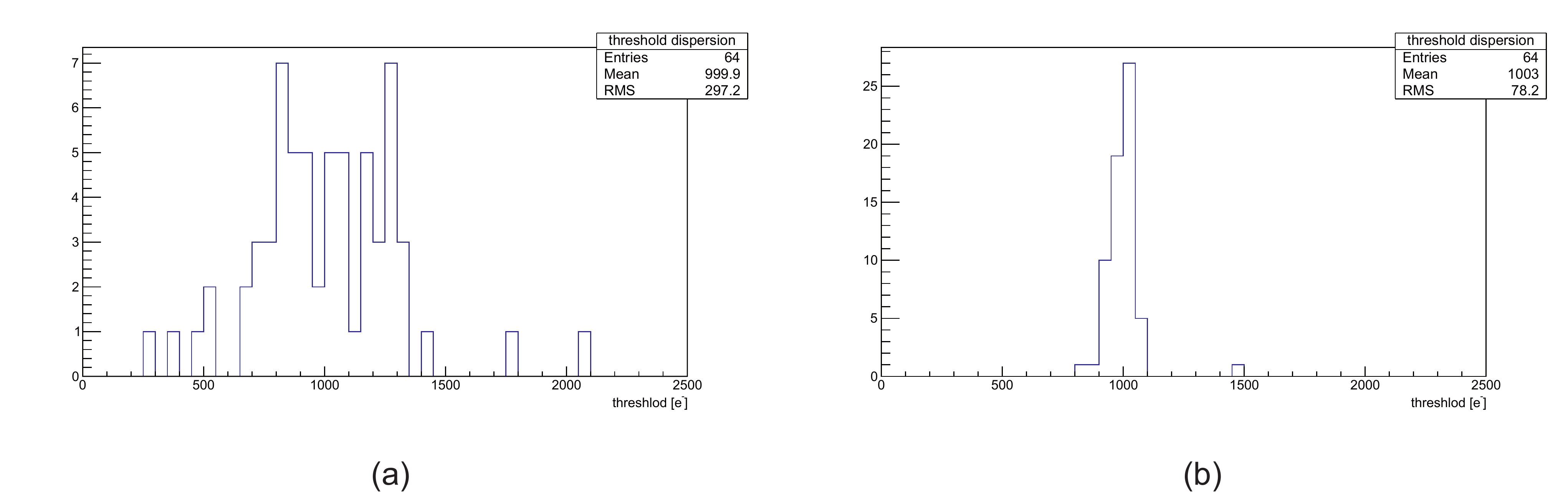}
\end{minipage}

\begin{minipage}[t]{420pt}
\caption{Threshold dispersion of the DMAPS pixels of variant V5 before (a) and after threshold tuning (b).}
\label{fig:thr_disp_v5}
\end{minipage}
\end{figure}

Both variants of the TDAC have been proven to significantly reduce the threshold dispersion. The distributions of a 1 ke\textsuperscript{-} threshold setting before and after tuning 
of pixels of variants V2 is shown in figure~\ref{fig:thr_disp_v2}~(a,b) and threshold dispersion of variant V5 is shown in figure~\ref{fig:thr_disp_v5}~(a,b).   

\subsection{Cluster size measurement}  
\begin{figure}[h]
\centering
\begin{minipage}[c]{420pt}
\includegraphics[width=420pt]{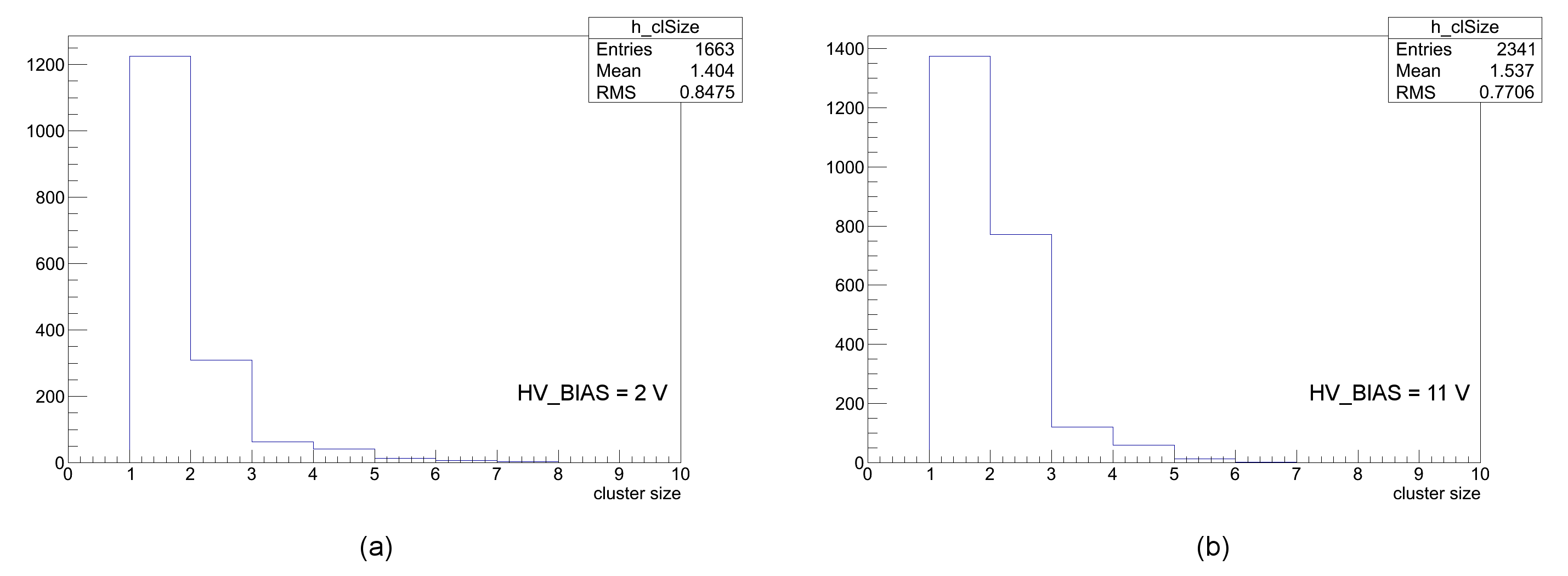}
\end{minipage}

\begin{minipage}[t]{420pt}
\caption{Distribution of the cluster size measured with the DMAPS matrix of variant V2 biased with 2~V (a) and 11~V (b).}
\label{fig:clusters}
\end{minipage}
\end{figure}

In a fine-pitch pixel sensor, the signal charge originating from an ionizing particle diffuses on its path to the electrode. 
The signal charge can be collected by a cluster of neighboring pixels, the size of which depends on the depleted thickness
of the traversed sensor.  
The cluster size is measured using the DMAPS pixels upon radiation from a \textsuperscript{90}Sr radioactive source.
The~cluster analysis has been performed with the DMAPS pixel array of V2 at several sensor bias voltages (HV\textunderscore BIAS). 
The detection threshold has been adjusted to~1~ke\textsuperscript{-} and has been equalized over the~matrix.
Distributions of the cluster size measured at bias voltages of 2 V and 11 V are shown in figure~\ref{fig:clusters}. 
By comparing these distributions, we can see that at 2 V bias voltage, 29\% fewer events are recorded than with a 
bias voltage setting of 11 V. Events from \textsuperscript{90}Sr are predominantly single pixel clusters.
However, double pixel clusters become more pronounced at 11 V than at 2 V. This can be explained by assuming that 
at a higher bias voltage a larger sensor volume is depleted and the total charge spreading by diffusion becomes larger. 
Consequently, a larger cluster size results.

\begin{figure}[h]
\centering
\begin{minipage}[c]{300pt}
\includegraphics[width=300pt]{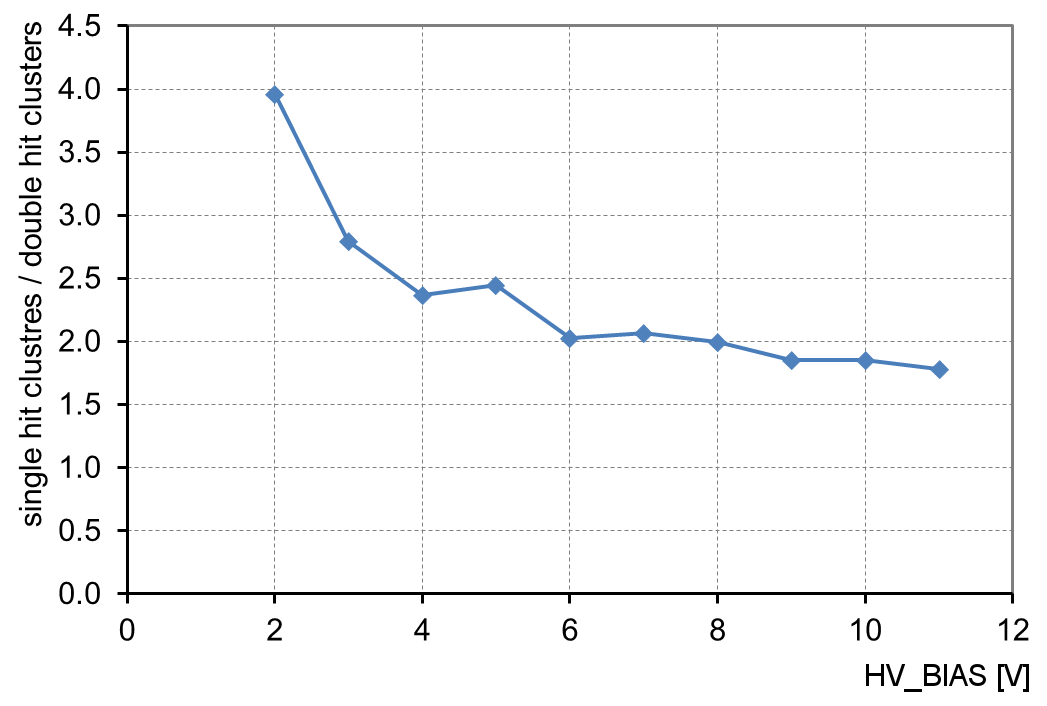}
\end{minipage}

\begin{minipage}[t]{300pt}
\caption{Ratio of the single pixel and double pixel clusters.}
\label{fig:shc_dhc}
\end{minipage}
\end{figure}

Figure~\ref{fig:shc_dhc} shows the ratio of single to double pixel clusters as a function of the sensor bias voltage.
This ratio decreases with increasing sensor bias voltage and saturates at a voltage of about 6~V, while the event rate 
does not increase beyond this point. At a voltage of 6~V and above the sensor does not collect any more charge and the
cluster size ratio remains constant. This, we conclude, indicates full depletion of the sensor.

\subsection{Gain determination with \textsuperscript{55}Fe}

The response of the pixel FE electronics should not (in ideal case) depend on whether the signal comes from~an~external charge injection circuit
or from the collection electrode. To verify this, an~independent measurement of the~gain of the pixel FE electronics using an    
\textsuperscript{55}Fe source has been performed. This source emits gamma rays with a characteristic peak in the energy spectrum
at 5.9~keV, which translates into a~signal of 1640 electrons. The energy of this peak has been measured by each pixel and the gain of 
the FE electronics has been determined. The same procedure has been completed with an external charge injection of signal of 1640 electrons.

\begin{figure}[h]
\centering
\begin{minipage}[c]{300pt}
\includegraphics[width=300pt]{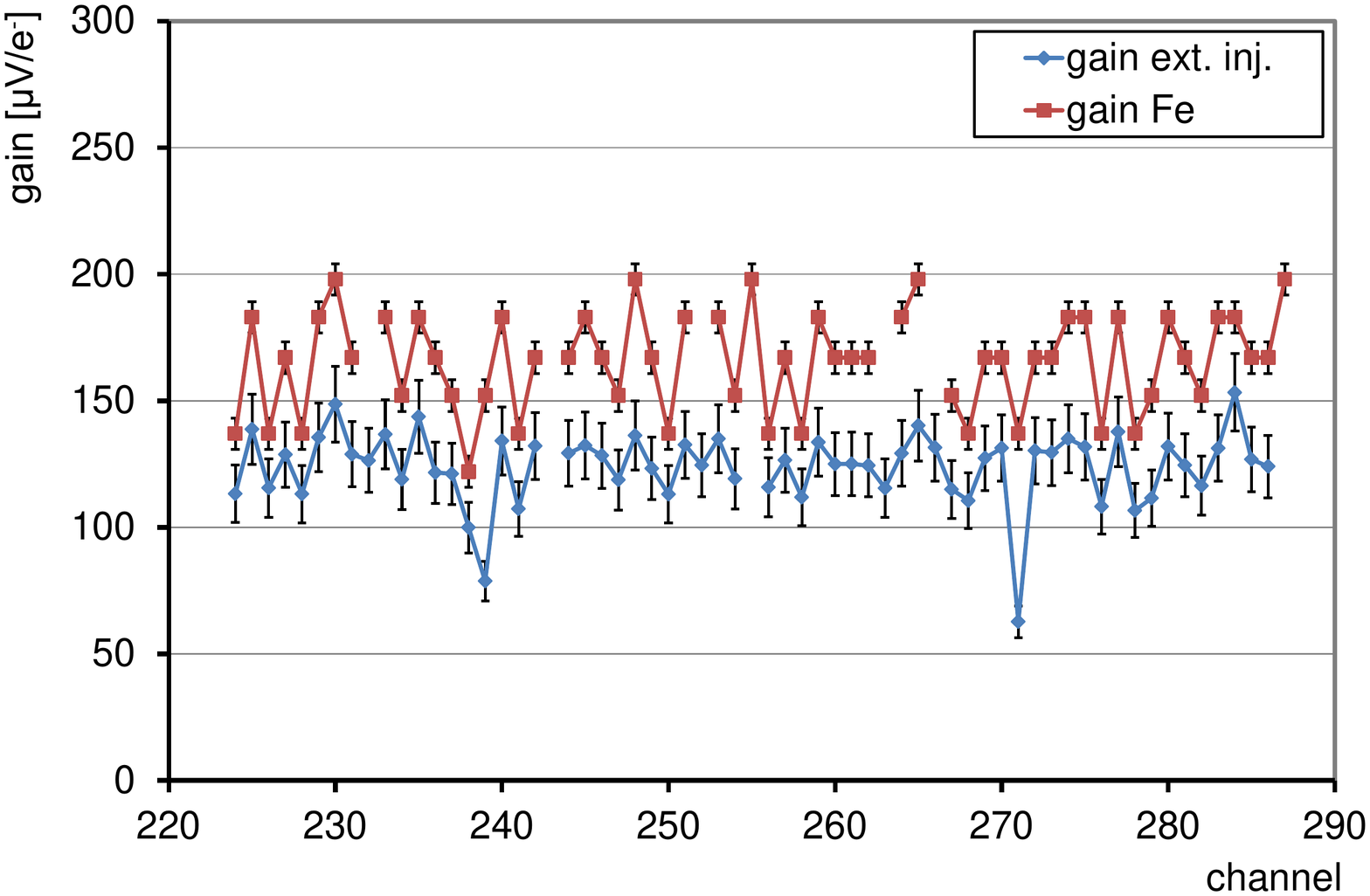}
\end{minipage}

\begin{minipage}[t]{300pt}
\caption{Gain of the FE electronics of each pixel of the matrix V2 determined independently by an external charge injection 
(\(\langle\)gain\(\rangle\)=124 \(\mu\)V/e\textsuperscript{-}) and by radioactive source \textsuperscript{55}Fe 
(\(\langle\)gain\(\rangle\)=166~\(\mu\)V/e\textsuperscript{-}).}
\label{fig:gain_Fe}
\end{minipage}
\end{figure}

Gain of each pixel of V2 determined by both methods (\textsuperscript{55}Fe irradiation and charge injection) is shown in figure~\ref{fig:gain_Fe}. 
Two effects can be seen: 1. the gain fluctuations determined by \textsuperscript{55}Fe are strongly correlated with those seen when the signal was provided by the external 
charge inejction. This result indicates that the gain fluctuations between pixels
are predominantly caused by the gain fluctuations of the FE electronics rather than by a non-uniform charge collection efficiency~of~the~collecting electrodes.~2.~the mean gain 
determined by \textsuperscript{55}Fe is on average by 34\% higher than the mean gain determined by the~charge 
injection. This effect has been studied in detail by design simulations and is most likely caused by the parasitic capacitance
of the charge injection circuit and that~of~the~capacitor coupling the collection electrode and electronics. 


\section{Radiation tests}\label{sec:radiationf}

\begin{figure}[h]
\centering
\begin{minipage}[c]{280pt}
\includegraphics[width=280pt]{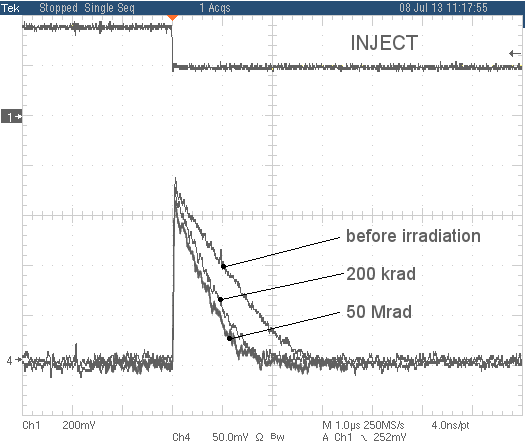}
\end{minipage}

\begin{minipage}[t]{280pt}
\caption{Signal pulse at the output of the CSA (V2) shortens after irradiation.}
\label{fig:pulse_shortening}
\end{minipage}
\end{figure}

\begin{figure}[h]
\centering
\begin{minipage}[c]{420pt}
\includegraphics[width=420pt]{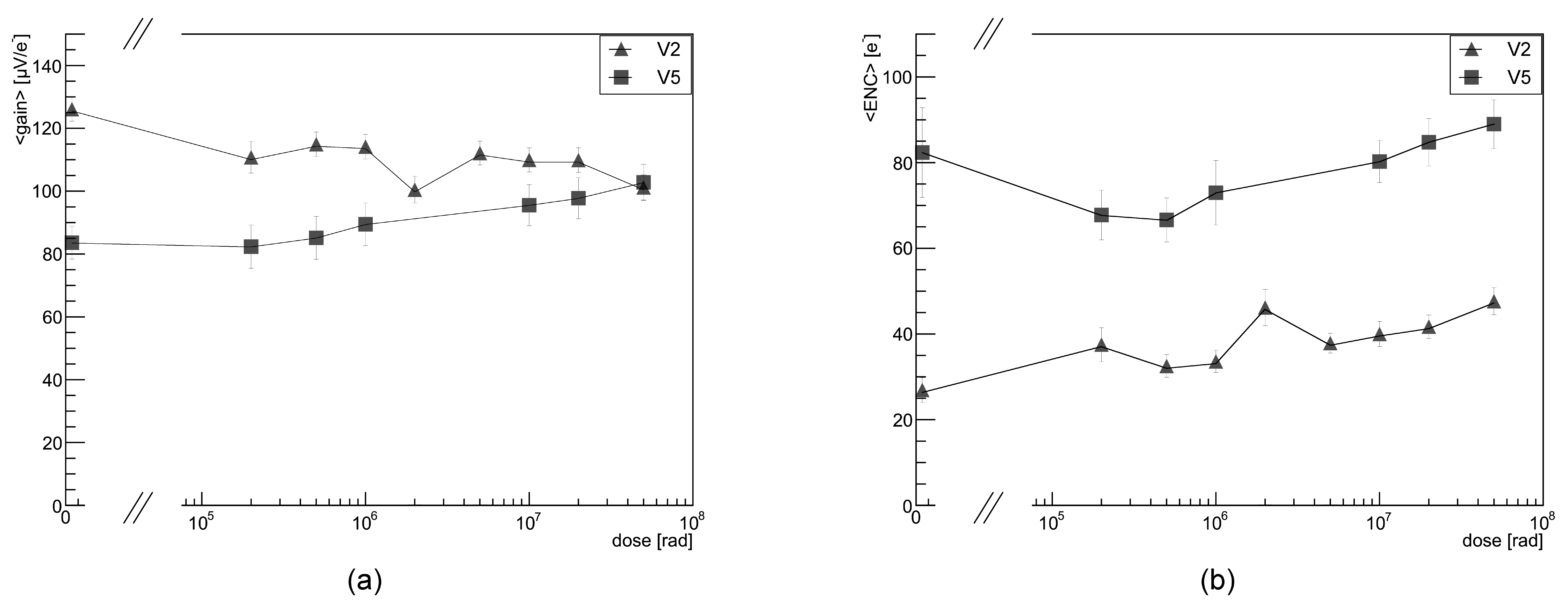}
\end{minipage}

\begin{minipage}[t]{420pt}
\caption{Gain (a) and noise (b) of the DMAPS pixels as a function of X-ray radiation dose. Error bars have been reduced by factor of~2.}
\label{fig:irrad}
\end{minipage}
\end{figure}

Radiation tolerance of the test chip EPCB01 has been investigated with irradiation from an X-ray tube with end-point energy of 60~keV.~The~irradiation has been performed within
several steps achieving a total ionizing dose of 50 Mrad. After each irradiation period, the chip has been annealed for 100 minutes
at 80 \textsuperscript{o}C. Several tests of the analog and digital part of the pixels have been performed after each irradiation period.
Radiation induced effects have only been observed in the~analog part of the FE electronics. The most sensitive node of the analog part of the DMAPS pixel is the NMOS feedback
transistor in the CSA. Radiation induced shift of the~threshold voltage of the feedback transistor changes the discharge time of the CSA. The~greatest difference of the~discharge 
time has been observed between an unirradiated state and after the first irradiation period (200 krad), then the discharge time changed only insignificantly
up to 50~Mrad as can be seen in figure~\ref{fig:pulse_shortening}. The X-ray irradiation has an overall impact on the FE electronics in terms of changes of gain and noise level of the DMAPS pixels. 
The dependence of these parameters on the~level of irradiation is shown in figure~\ref{fig:irrad}~(a,b).
The digital part of the~DMAPS pixels has been tested between irradiation periods  by writing the test data patterns in the configuration
shift register and reading them back. No difference has been observed in the data patterns passing the~configuration register of the 
irradiated EPCB01. \\

\begin{figure}[h]
\centering
\begin{minipage}[c]{420pt}
\includegraphics[width=420pt]{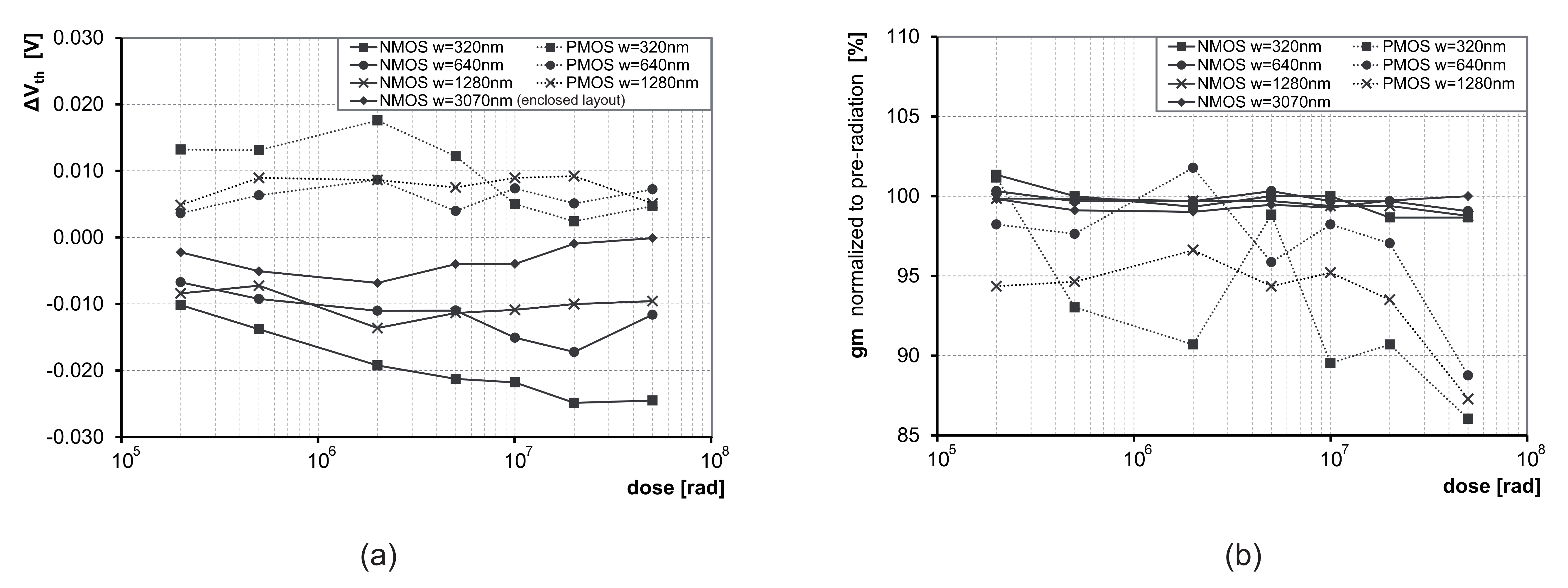}
\end{minipage}

\begin{minipage}[t]{420pt}
\caption{Threshold voltage shift $\Delta V_{\mathrm{th}}$ (a) and transconductance (b) of MOSFET transistors
as a function of total ionizing dose. Transistors differ by channel width $w$ while channel length remains constant (150~nm).}
\label{fig:radTran}
\end{minipage}
\end{figure}

An additional radiation studies have been performed on an array of individual NMOS and PMOS transistors which is part of EPCB01.
All transistors in the array have constant channel length of 150~nm but variable channel width. The NMOS transistor with channel 
width of 3070~nm has enclosed layout geometry while all other transistors have standard layout. Radiation induced shift of threshold voltage 
has been observed as well as degradation of transconductance of the~transistors. Both results are shown 
in figure~\ref{fig:radTran}~(a,b). Threshold voltage of both transistor types shifts by less than 30~mV within the range of radiation dose. 
Maximum transconductance of the NMOS transistors degrades by less than 2\(\%\) and in case of PMOS transistors transconductance degrades
by less than 15\(\%\). In general, the radiation effects are more significant in small channel width transistors.\\


\section{Summary}\label{sec:summary}
A novel concept of Depleted Monolithic Active Pixel Sensors (DMAPS) has been introduced. DMAPS pixels integrate a complex "hybrid pixel like" 
CMOS electronics and simultaneously benefit from a fully depleted sensor within the same substrate. A commercial CMOS process has been used for 
fabrication of a prototype DMAPS chip~-~EPCB01. First tests indicate good functionality of this detector concept, while there 
is ample room for improvement of the~FE electronics. The~best performance has been achieved with an AC coupled diode biased collection electrode connected 
to the time-continuous FE electronics. 
The gain is $\sim$100~\(\mu\)V/e\textsuperscript{-}, the noise is $\sim$30 e\textsuperscript{-} with~a 
discharge time of about 1~\(\mu\)s. The threshold 
dispersion after equalization is 135 e\textsuperscript{-}. The RTS noise has been observed in all variants of the DMAPS pixels and most likely emerges from 
fluctuations of current of close to minimum size transistors used in the FE electronics. The depletion voltage of the high resistivity 
substrate has been determined by the means of cluster size saturation to be approximatelly 6 V. Small radiation effects have been observed in the performance of the 
analog FE electronics after absorbing the X-ray radiation dose of 50~Mrad. More measurements need to be done to fully understand the DMAPS pixels.
In particular, laser scans are needed to determine the~position dependence of the charge collection efficiency and more radiation test are needed to
learn about the radiation effects in the charge collecting part. However, important lessons have already been learned during the design and testing of the EPCB01 and they 
will be addressed in the~next generation of the DMAPS test chip.

\section{Acknowledgement}\label{sec:ack}
Authors would like to thank to E. Marchesi and M. Popp from ESPROS Photonics AG for providing information about their technology. 
Another thanks belong to C.~Marinas from Bonn for organization of radiation tests and to T. M\"uller, H. J. Simonis and S. Heindl from Karlsruhe Institute of Technology 
for allowing us to perform X-ray irradiation of our chip at their laboratory. At last but not least we would like to thank to W. Dietsche and W. Ockenfels for wire-bonding 
the~test-chips.


\vspace{1cm}
\bibliography{DMAPS}
\bibliographystyle{unsrt}

\end{document}